\documentclass[twocolumn,aps,showpacs,prl,amsmath]{revtex4-1}

\usepackage{amssymb}
\usepackage{graphicx}
\usepackage{lipsum}  
\newcommand{\bd}{\boldsymbol{\delta}}
\newcommand{\bj}{\mathbf{j}}
\newcommand{\bt}{\mathbf{t}}
\newcommand{\br}{\mathbf{r}} 

\newcommand{\bk}{\mathbf{k}}

\newcommand{\bg}{\mathbf{g}} 
\newcommand{\bA}{\mathbf{A}}

\newcommand{\bR}{\mathbf{R}}

\begin{document}

\title{Comment on``Orientation dependence of the optical spectra \\ in
graphene at high frequencies"} 
\author{V. Hung Nguyen$^{1,3}$ and Huy-Viet Nguyen$^{2,3}$}
\address{$^1$Institute of Condensed Matter and Nanosciences, Universit\'{e}
catholique de Louvain, Chemin des \'{e}toiles 8, B-1348 Louvain-la-Neuve,
Belgium}
\address{$^2$Institute of Research and Development, Duy Tan University, K7/25
Quang Trung, Da Nang, Viet Nam}
\affiliation{$^3$Institute of Physics, Vietnam Academy of
Science and Technology, 18 Hoang Quoc Viet, Cau Giay, Ha Noi, Viet Nam}
%
%
%
\begin{abstract}
Zhang \textit{et al.} reported in [Phys. Rev. B \textbf{77}, 241402(R) (2008)]
a theoretical study of the optical spectra of monolayer graphene employing
the Kubo formula within a tight-binding model. Their calculations predicted that
at high frequencies the optical conductivity of graphene becomes strongly
anisotropic. In particular, at frequencies comparable to the energy separation
of the upper and lower bands at the $\Gamma$-point, the optical conductivity
is strongly suppressed if the field polarization is along the zigzag direction
while it is significantly high for the armchair one. We find that, unfortunately, this 
result is just a consequence of the incorrect determination of the current operator in 
$k$-space. Here, we present the standard scheme to obtain this operator correctly. 
As a result, we show that the optical conductivity of monolayer graphene is indeed 
isotropic, which is consistent with the results of other (both theoretical and experimental)
studies in the literature.
\end{abstract}

\maketitle

Current operator is the key ingredient in the calculation of optical spectra from the 
Kubo formula. For a periodic system, it is most convenient to
evaluate this quantity using the $k$-space representation of the Hamiltonian
and the corresponding Bloch's wave functions. For graphene, the use of
$p_z$-orbital tight-binding models has been shown to provide a good
description of electronic states for many purposes. However, for monolayer graphene where 
\textit{C}-atoms are arranged in a honeycomb lattice with the unit cell 
containing two atoms, there are two tight-binding bases widely used in the literature. 
Because of this, care must be taken in order to
avoid the use of inappropriate forms of certain operators which may
lead to erroneous physical predictions as already noted in
Ref.~\onlinecite{bena09}. In particular, the momentum operator, $\mathbf{p}$, for a
periodic system has often been determined by\cite{Blount}
\begin{equation}
\mathbf{p} = \frac{m_0}{\hbar}\nabla_{\bk}H(\bk),
\label{Eq:p}
\end{equation}
where $m_0$ is the free electron mass and $\nabla_{\bk}H(\mathbf{k})$ is the
gradient of $k$-dependence representation of the Hamiltonian. Obviously, the
form of $H(\bk)$ is not unique in that it depends on the tight-binding
basis used\cite{bena09} or even on the choice of unit cell.\cite{Gundra} Hence, 
one can get different results computing $\mathbf{p}$ from Eq.~(\ref{Eq:p}). 
One way to avoid this issue and to achieve correctly the $k$-dependence representation 
of operators is to use their original definitions and then represent them 
in $k$-space, as exemplified by the calculation of current operator in graphene below.

We start from the standard formula of the current operator within
an independent electron approximation:
\begin{eqnarray}
\mathbf{j} &=& e\mathbf{v} = \frac{e}{m_0}\mathbf{p}\\
\mathbf{p} &=& \frac{m_0}{i\hbar} \left[ \br, H \right] 
\end{eqnarray}

The tight-binding Hamiltonian, in the first nearest-neighbor approximation, is written as
\begin{equation}
 H = -t\sum\limits_{\bR,\bd} c^{\dagger}_{\bR}c_{\bR+\bd},
 \label{Eq:Hreal}
\end{equation}
where $c^{\dagger}_{\bR}$ and $c_{\bR+\bd}$ are creation and annihilation
operators, respectively, for $p_z$-electrons located at the site $\bR$ and its first
nearest neighbors $\bR+\bd$. Then, the current operator is
\begin{equation}
\mathbf{j} = -i\frac{et}{\hbar} \sum\limits_{\bR,\bd} 
\bd c^{\dagger}_{\bR}c^{}_{\bR+\bd}. 
\label{Eq:j}
\end{equation}
Actually, this expression can also be obtained
following another scheme by introducing the vector potential $\bA$ in the
tight-binding Hamiltonian in Eq.~(\ref{Eq:Hreal}). Using the Peierls
substitution:
\begin{eqnarray}
t_{ij} \rightarrow t_{ij} e^{-i\frac{e}{\hbar} \bA{\cdot}(\br_j - \br_i)},
\end{eqnarray}
the current operator is then determined by
\begin{equation}
\mathbf{j} = - \left. \frac{\partial H[\bA]}{\partial \bA} \right|_{\bA
\rightarrow 0},
\label{Peisub}
\end{equation}
which leads to an identical formula as in Eq.~(\ref{Eq:j}).

In monolayer graphene with two atoms in its primitive cell, these operators
can be rewritten in the following forms:
\begin{eqnarray}
H &=& \sum\limits_{nm}\left(H_{nm} + H_{nm,\pm\bt_1} + H_{nm,\pm\bt_2}\right)\\
H_{nm} &=& -t(a^{\dagger}_{\bR_{nm}} b_{\bR_{nm} - \bd_3} +
b^{\dagger}_{\bR_{nm} - \bd_3}a_{\bR_{nm}}) \nonumber \\
H_{nm,\bt_{1,2}} &=& -tb^{\dagger}_{\bR_{nm} - \bd_3} 
a_{\bR_{nm} + \bt_{1,2}} \nonumber \\
H_{nm,-\bt_{1,2}} &=& -ta^{\dagger}_{\bR_{nm}} b_{\bR_{nm} - \bd_3 - \bt_{1,2}},
\nonumber
\end{eqnarray} 
and
\begin{eqnarray}
\bj &=& \sum\limits_{nm} \left(\bj_{nm} + \bj_{nm,\pm\bt_1} +
\bj_{nm,\pm\bt_2}\right) \\
\bj_{nm} &=& i\frac{et}{\hbar} \bd_3 (a^{\dagger}_{\bR_{nm}}
b_{\bR_{nm}-\bd_3} - b^{\dagger}_{\bR_{nm}-\bd_3}a_{\bR_{nm}})
\nonumber \\
\bj_{nm,\bt_{1,2}} &=& -i\frac{et}{\hbar} \bd_{1,2}
b^{\dagger}_{\bR_{nm}-\bd_3} a_{\bR_{nm}+\bt_{1,2}}
\nonumber \\
\bj_{nm,-\bt_{1,2}} &=& i\frac{et}{\hbar} \bd_{1,2}
a^{\dagger}_{\bR_{nm}}b_{\bR_{nm}-\bd_3-\bt_{1,2}}, \nonumber
\end{eqnarray} 
where $\bR_{nm} = n\bt_1 + m\bt_2$, $\bd_3 = -(\bt_1 + \bt_2)/3$, $\bd_{1,2} = \bt_{1,2} + \bd_3$,
and $\bt_{1,2}$ are two primitive lattice vectors (see Fig.~1). Here we have
distinguished two types of creation and annihilation operators: $a^{\dagger}, a$
for atoms in the sublatice A and $b^{\dagger}, b$ for atoms in the sublatice B.
\begin{figure}[!t]
\centering
\includegraphics[width = 0.3\textwidth]{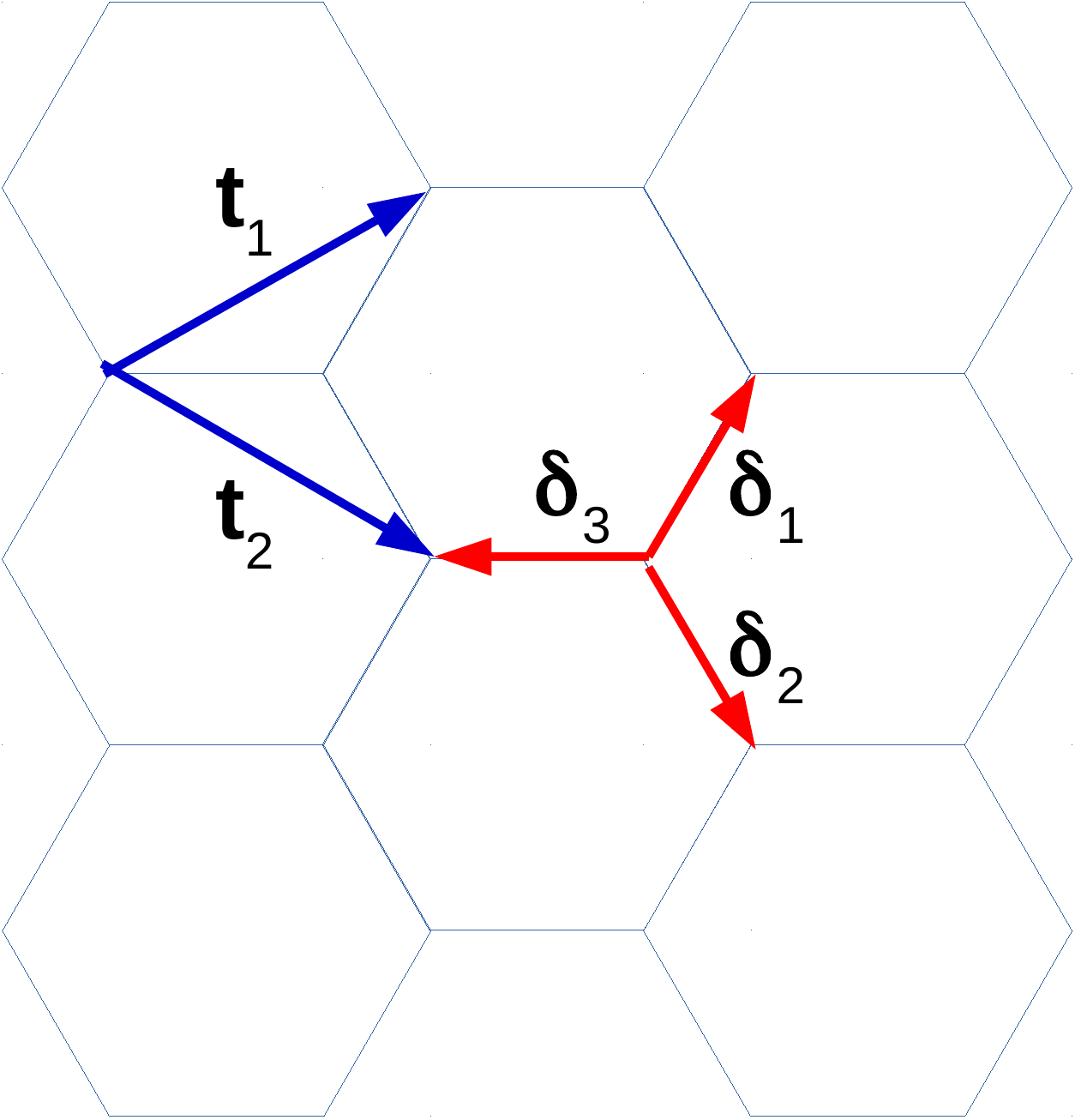}
\caption{Schematic of graphene lattice with the conventions for vectors
$\bt_{1,2}$ and $\bd_{1,2,3}$ used in the text.}
\label{fig_sim0}
\end{figure}

As mentioned in Ref.~\onlinecite{bena09}, there are two tight binding bases
most commonly used to describe graphene in the literature. Accordingly, there
are two forms of the Fourier transformation for $a$ and $b$ operators: 
\begin{eqnarray}
\label{Eq:FT11}
a_{nm} &=& \frac{1}{\sqrt{N_{cell}}} \sum\limits_{\bk}
a_{\bk} e^{i\bk{\cdot}(\bR_{nm}+\br_a)} \\
\label{Eq:FT12}
b_{nm} &=& \frac{1}{\sqrt{N_{cell}}} \sum\limits_{\bk}
b_{\bk} e^{i\bk{\cdot}(\bR_{nm}+\br_b)}
\end{eqnarray}
and
\begin{eqnarray}
\label{Eq:FT21}
a_{nm} &=& \frac{1}{\sqrt{N_{cell}}} \sum\limits_{\bk}
\tilde{a}_{\bk} e^{i\bk{\cdot}\bR_{nm}} \\
\label{Eq:FT22}
b_{nm} &=& \frac{1}{\sqrt{N_{cell}}} \sum\limits_{\bk}
\tilde{b}_{\bk} e^{i\bk{\cdot}\bR_{nm}},
\end{eqnarray}
with $N_{cell}$ being the number of periodic (primitive) cells. With these two
Fourier transformations, the Hamiltonian (8) is respectively rewritten in two
different $k$-dependent forms:
\begin{eqnarray}
\label{Eq:Hk1}
H &=& -t \sum\limits_{\bk} \left[h(\bk)b_{\bk}^{\dagger}a_{\bk} + h.c.\right] \\
\label{Eq:Hk2}
H &=& -t \sum\limits_{\bk} \left[\tilde{h}(\bk)
\tilde{b}_{\bk}^{\dagger} \tilde{a}_{\bk} + h.c.\right]
\end{eqnarray}
where $h(\bk) = e^{i\bk{\cdot}\bd_1} + e^{i\bk{\cdot}\bd_2} +
e^{i\bk{\cdot}\bd_3}$ and $\tilde{h}(\bk) = e^{i\bk{\cdot}\bt_1} +
e^{i\bk{\cdot}\bt_2} + 1 \equiv e^{-i\bk{\cdot}\bd_3} h(\bk)$. Similarly, the
current operator (9) is respectively rewritten in two different forms: 
\begin{eqnarray}
\label{Eq:jk1}
\bj &=& -\frac{et}{\hbar} \sum\limits_{\bk}
\left[\bg(\bk) b_{\bk}^{\dagger} a_{\bk} + h.c.\right] \\
\label{Eq:jk2}
\bj &=& -\frac{et}{\hbar} \sum\limits_{\bk}
\left[\tilde{\bg}(\bk) \tilde{b}_{\bk}^{\dagger} \tilde{a}_{\bk} + h.c.\right]
\end{eqnarray}
where $\bg(\bk) = i(\bd_1 e^{i\bk{\cdot}\bd_1} + \bd_2 e^{i\bk{\cdot}\bd_2}
+ \bd_3 e^{i\bk{\cdot}\bd_3})$ and $\tilde{\bg}(\bk) =
i(\bd_1 e^{i\bk{\cdot}\bt_1} + \bd_2 e^{i\bk{\cdot}\bt_2} + \bd_3) \equiv 
e^{-i\bk{\cdot}\bd_3} \bg(\bk)$.
\begin{figure}[!b]
\centering
\includegraphics[width = 0.48\textwidth]{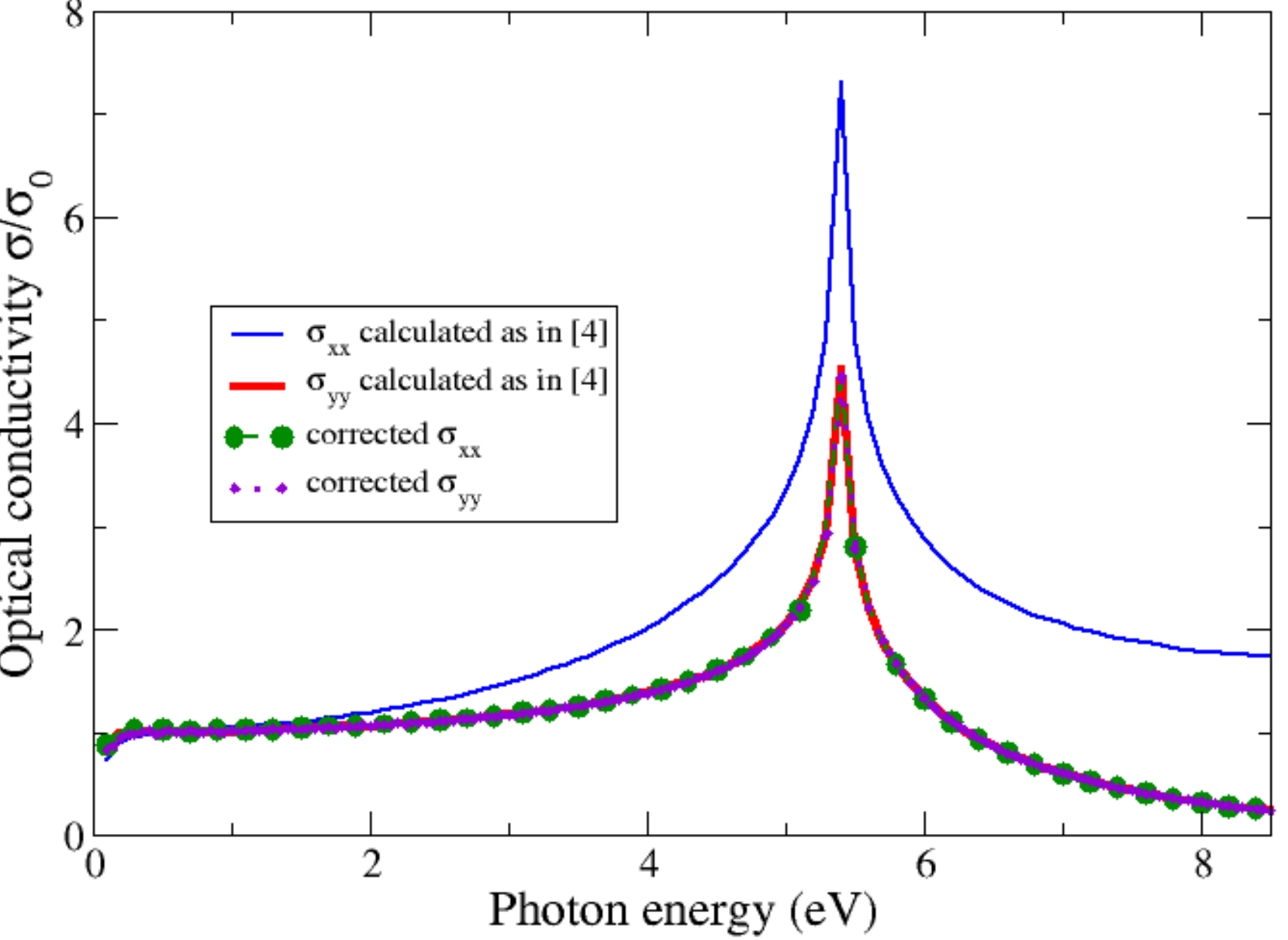}
\caption{Corrected optical conductivity in comparison with the calculations of
Zhang \textit{et al.} in Ref.~\onlinecite{zhan08} ($\sigma_0 = e^2/4\hbar$).}
\label{fig2}
\end{figure}

\begin{figure*}[!t]
\centering
\includegraphics[width = 0.90\textwidth]{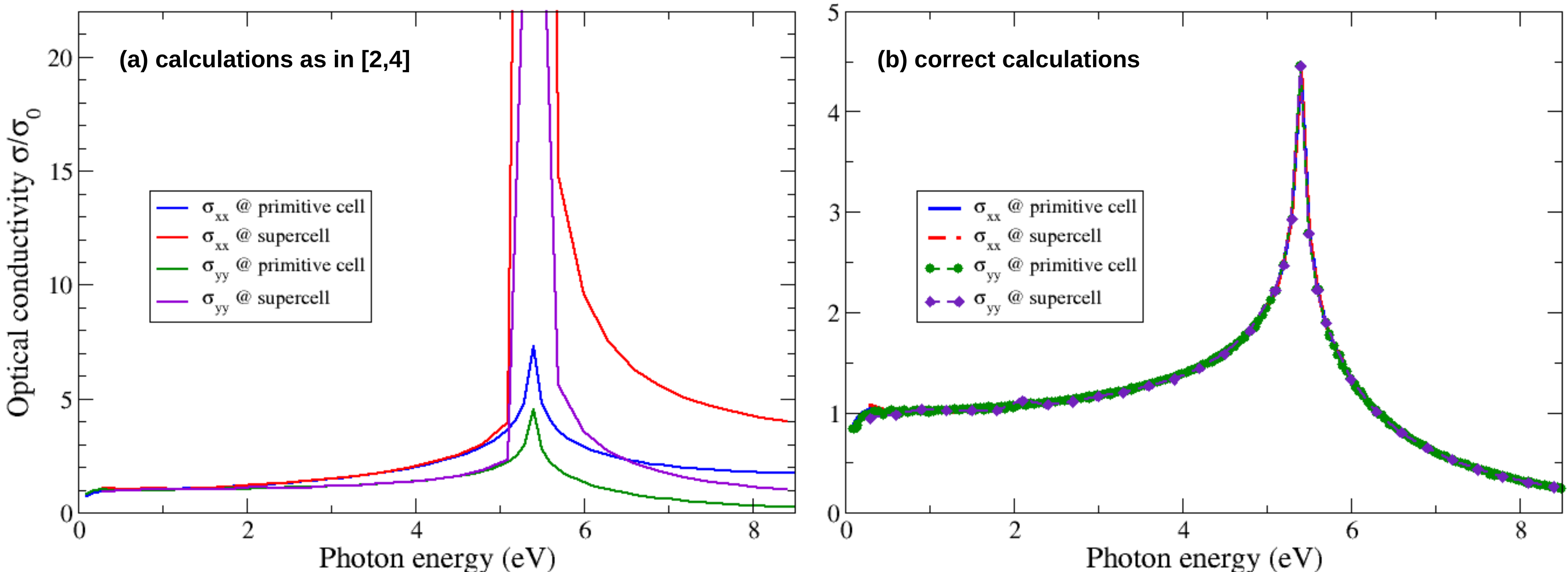}
\caption{(a) Incorrect unit-cell dependence of optical conductivity obtained
using the scheme as in Refs.~\onlinecite{Gundra,zhan08}, in comparison with (b)
the correct ones obtained using the scheme presented in this work ($\sigma_0 =
e^2/4\hbar$).}
\label{fig3}
\end{figure*}
Now, the Hamiltonians in Eqs.~(\ref{Eq:Hk1}-\ref{Eq:Hk2}) are solved to compute
their eigenvalues and eigenfunctions. In both cases, one obtains the same eigenvalue
$\varepsilon_s(\bk) = st\left| h(\bk)\right| \equiv st| \tilde{h}(\bk)|$ with 
$s = \pm 1$ for the conduction/valence bands, respectively. 
The corresponding eigenfunction has the form $\psi_s(\bk) =
\frac{1}{\sqrt{2}}\left( {\begin{array}{*{20}{c}} {-se^{i\theta_\bk}}\\
{1} \end{array}} \right)$ where $e^{i\theta_\bk} = h(\bk)/\left|
h(\bk)\right|$ and $\tilde{h}(\bk)/\left|\tilde{h}(\bk)\right|$,
respectively, for the Hamiltonians (\ref{Eq:Hk1}) and (\ref{Eq:Hk2}). 
Using these eigenfunctions, one can make a
transformation to recast the Hamiltonian (\ref{Eq:Hk1}-\ref{Eq:Hk2}) and
the current operator (\ref{Eq:jk1}-\ref{Eq:jk2}) to the following forms:
\begin{eqnarray}
H &=& \sum\limits_{\bk,s} \varepsilon_s(\bk)c_{\bk,s}^{\dagger}
c_{\bk,s},\\
\bj &=& \frac{et}{\hbar} \sum\limits_{\bk,s} 
s\frac{\text{Re}\left\{h^{\ast}(\bk)\bg(\bk)\right\}}{\left|h(\bk)\right|}
c_{\bk,s}^{\dagger} c_{\bk,s} \nonumber \\
&& +i\frac{et}{\hbar}\sum\limits_{\bk,s} 
s\frac{\text{Im}\left\{ h^{\ast}(\bk)\bg(\bk)\right\}}{\left|h(\bk)\right|}
c_{\bk,s}^+ c_{\bk,-s}.
\end{eqnarray}
Note that, as for the case of the Hamiltonian operator (Eq.~(18)),  one obtains the 
same formula for the current operator (Eq. (19)) regardless of the Fourier 
transformations used because 
$h^{\ast}(\bk)\bg(\bk)\equiv\tilde{h}^{\ast}(\bk)\tilde\bg(\bk)$. This is 
consistent with the remarks in Ref.~\onlinecite{bena09} that if all operators are 
represented in the same basis, the expectation value of observable quantities is 
independent on the tight-binding basis.
\begin{widetext}
\noindent Since
\begin{eqnarray*}
h^{\ast}(\bk)\bg(\bk) &=& i[\bd_1 + \bd_2 + \bd_3 + \bd_1e^{ik_yr_0}
+ \bd_2 e^{-ik_yr_0} 
+(\bd_1e^{i\frac{k_yr_0}{2}} + \bd_2 e^{-i\frac{k_yr_0}{2}})
e^{i\frac{k_xr_0\sqrt{3}}{2}} 
+2\bd_3 \cos(\frac{k_yr_0}{2})e^{-i\frac{k_xr_0\sqrt{3}}{2}} ],
\end{eqnarray*}
after some straightforward manipulations one ends up with the following 
expressions for the current operator
\begin{eqnarray}
j_x &=& -2e\text{v}_F\sum\limits_{\bk,s}
s\frac{\cos(\frac{k_yr_0}{2})\sin(\frac{k_xr_0\sqrt{3 }}{2})}{\left|
h(\bk)\right|}c_{\bk,s}^{\dagger} c_{\bk,s}
+i\frac{2e\text{v}_F}{3}\sum\limits_{\bk,s} s\frac{\cos(k_yr_0) -
\cos(\frac{k_yr_0}{2})\cos(\frac{k_xr_0\sqrt{3}}{2}) }{\left|
h(\bk)\right|}c_{\bk,s}^{\dagger} c_{\bk,-s}, \nonumber \\
j_y &=& -\frac{2e\text{v}_F}{\sqrt{3}} \sum\limits_{\bk,s}
s\frac{\sin(k_yr_0)+\sin(\frac{k_yr_0}{2})\text{cos}(\frac{k_xr_0\sqrt{3}}{2})}
{\left| h(\bk)\right|}c_{\bk,s}^{\dagger} c_{\bk,s}
-i\frac{2e\text{v}_F}{\sqrt{3}}\sum\limits_{\bk,s}
s\frac{\sin(\frac{k_yr_0}{2})\sin(\frac{k_xr_0\sqrt{3}}{2}) }
{\left|h(\bk)\right|}c_{\bk,s}^{\dagger} c_{\bk,-s}, \nonumber
\end{eqnarray}
where $r_0$ denotes the $C-C$ bond length in graphene.
\end{widetext}

Compared to the expressions for current operator presented by Zhang 
\textit{et al.} in Ref.~\onlinecite{zhan08}, the $j_y$-component obtained here is identical to 
theirs, but it is not the case for $j_x$-component. We note that even though the 
$j_x$- and $j_y$-components have different $k$-dependence, the integral over the 
whole Brillouin zone in the Kubo formula \cite{stau08,pell10} gives the same optical 
conductivities $\sigma_{xx}$ and $\sigma_{yy}$ as displayed in Fig.~\ref{fig2},
i.e. \textit{the optical spectra of graphene is indeed isotropic}, which is at
variance with the anisotropic behavior shown in calculations by Zhang \textit{et
al.}. Additionally, the value of optical conductivity in the low frequency limit
reported in Ref.~\onlinecite{zhan08} is $e^2/2\hbar$ which is twice the
well-known value of $\sigma_0 = e^2/4\hbar$ for monolayer 
graphene\cite{fmak11}. Note that our obtained results are in good agreement with 
those reported (both theoretically with different methods \cite{stau08,yang09,pell10,novk16} and 
experimentally \cite{fmak11}) in the literature. The anisotropy of optical spectra 
can be achieved only if the symmetry properties of graphene lattice are broken, e.g., by 
strain effects as demonstrated in Ref. \onlinecite{pell10}.

In Ref.~\onlinecite{zhan08}, the authors provided no information on how the current
operator was actually calculated. However, one could reproduce their
expressions for $j_{x,y}$ when using the formula $j_{\mu} =
\frac{e}{\hbar}\frac{\partial H}{\partial k_{\mu}}$ -- indeed used by Zhang
\textit{et al.} in other studies \cite{nano09,phrl09} -- with the 
Hamiltonian in Eq. (1) of Ref.~\onlinecite{zhan08} (i.e., Eq. (15) here).
Obviously, $j_\mu$ determined in this way is not correct because 
$\partial \tilde{h}(\bk)/\partial k_\mu$ is not identical to $\tilde{g}_\mu(\bk)$. 
The expression for $j_y$-component in Ref.~\onlinecite{zhan08} is fortuitously correct
just because the $y$-component of vectors $\bt_{1,2}$ are identical to that of
vectors $\bd_{1,2}$, respectively, while $\bd_{3y} = 0$. Hence, we speculate
that the use of the formula $j_{\mu} = \frac{e}{\hbar}\frac{\partial H}{\partial
k_{\mu}}$ with the Hamiltonian in Eq.~(15) is the origin of the erroneous results 
obtained by Zhang \textit{et al.}. We would like to note additionally that the use of
this incorrect determination of the
current operator also results in the unit cell dependence of optical matrix
elements (see Fig.~\ref{fig3}(a) and in Ref.~\onlinecite{Gundra}). 
Basically, the calculations using supercells lead to the band folding, compared to that of primitive cell.
Using the incorrect formulas of current operators can allow for unphysical transitions between the folding bands and hence gives wrong results at high energies (see Fig.~\ref{fig3}(a)). The authors in Ref.~\onlinecite{Gundra} tried to use group-theoretic
arguments to demonstrate that one would obtain incorrect results if the unit cell
chosen does not incorporate the symmetries of the bulk. Physically, these arguments do not sound
reasonable to us as any change in the unit cell only leads to a change in 
the matrix representation of operators and the 
calculated results should be, in principle, unchanged if the operators in the $k$-space 
are correctly determined. This is actually confirmed 
by the data presented in Fig.~\ref{fig3}(b) where our calculations were performed 
using the current operators determined from the original formula (5).

Thus, in order to achieve the correct formula for any operator in the 
$k$-space, we recommend that one should perform the Fourier transform with its 
original formula in real space. By this way, the obtained results should
depend neither on the tight-binding basis nor on the unit cell. This is because
the use of  another tight-binding basis or unit cell only leads to a change in 
the matrix representation of operators and hence the expectation
value of observable quantities should always be correctly achieved. However,
there are some specific quantities determined directly from the
Hamiltonian in the $k$-space and the phase of Bloch wave functions, e.g., the
Berry connection and Berry curvature. In such cases, it has been demonstrated
in Ref.~\onlinecite{fuch10} that only the Fourier transformation in
Eqs.~(\ref{Eq:FT11}-\ref{Eq:FT12}) gives the correct results. Similarly, the
current operator in Eq. (16) can be also obtained correctly by using the formula 
$j_\mu =\frac{e} {\hbar}\frac{\partial H}{\partial k_{\mu}}$ with the Hamiltonian 
in Eq. (14), i.e., $\partial h(\bk)/\partial k_\mu$ is indeed identical to $g_\mu(\bk)$.

To conclude, we have shown that the anisotropicity of the optical spectra
reported by Zhang \textit{et al.} in Ref.~\onlinecite{zhan08} is just a
consequence of the incorrect determination of the current operator in
the $k$-space. Starting from the original definition of the current operator in the real
space, we present a scheme to correctly obtain its formula in the $k$-space, regardless
of the tight binding basis as well as the choice of unit cell used in the
calculations. Our Comment thus emphasizes a simple but subtle and fundamental 
remark which will be of useful to researchers working with tight-binding
calculations, particularly, in graphene and its derivatives.
	
\textbf{\textit{Acknowledgment.}} We would like to thank Aur\'{e}lien Lherbier
and Jean-Christophe Charlier for their helpful comments and discussions. V.H.N.
acknowledges financial
support from the Fonds de la Recherche Scientifique de Belgique (F.R.S.-FNRS)
through the research project (N$^\circ$ T.1077.15), from the Commmunaut\'{e}
Wallonie-Bruxelles through the Action de Recherche Concert\'{e}e (ARC) on
Graphene Nano-electromechanics (N$^\circ$ 11/16-037) and from the European ICT
FET Flagship entitled "Graphene-Based Revolutions in ICT And Beyond" (N$^\circ$
604391). This research in Hanoi is funded by Vietnam's National Foundation for Science and Technology Development (NAFOSTED) under grant N$^\circ$ 103.01-2014.24.

\end{document}